\documentclass{PoS}
\usepackage{lineno}
\title{Calibration of the highly segmented SoLid antineutrino detector}

\ShortTitle{Calibration of the highly segmented SoLid antineutrino detector}

\author{\speaker{Luis Manzanillas}\thanks{On behalf of the SoLid  collaboration}\\
        LAL, Univ Paris-Sud, CNRS/IN2P3, Universit\'e Paris-Saclay, Orsay, France\\
        E-mail: \email{manzanillas@lal.in2p3.fr}}

\abstract{SoLid is a short baseline neutrino experiment, which is currently operating a 1.6 tons detector at the SCK$\bullet$CEN BR2 research reactor in Belgium. SoLid will address the study of the so called Reactor Antineutrino Anomaly (RAA), whose origin could be the existence of a light sterile neutrino state with a mass around the  eV scale. In addition, it will perform a new  measurement of the antineutrino energy spectrum produced by the $^{235}$U isotope, which will help in the understanding of the 5-MeV distortion observed in previous reactor antineutrino experiments.
SoLid leverages a novel technology, combining PVT cubes of 5$\times$5$\times$5 cm$^{3}$ dimensions and $^{6}$LiF:ZnS(Ag) screens of $\sim$250 $\mu$m thickness. To detect antineutrino interactions, signals are readout by a network of wavelength shifting fibers and SiPMs. The fine granularity (12800 cells) provides powerful tools to distinguish signal from background, but presents a challenge in ensuring homogeneous detector response and calibrating the energy scale and neutron detection efficiency. In this contribution the methods that have been developed for the calibration of such a segmented detector will be described. In addition, the calibration results are presented.}

\FullConference{
European Physical Society Conference on High Energy Physics - EPS-HEP2019 -\\
			10-17 July, 2019\\
			Ghent, Belgium}

\begin{document}

\section{Introduction}
Several anomalies in previous short baseline neutrino experiments has been evidenced in recent years. These anomalies are known as the Reactor Antineutrino Anomaly (RAA) and the Gallium Anomaly (GA), and both show discrepancies with respect to the expectations at the 3$\sigma$ level. Oscillations into a light sterile $\nu$ state at the eV scale could be at the origin of these anomalies \cite{Gariazzo:2017fdh,Dentler:2018sju}. Recent reactor neutrino experiments also observed a distortion in the $\overline{\nu}_e$ energy spectrum, which is known as the ``5-MeV bump'', whose origin remains unknown.  The consensus among the neutrino community is that new short baseline reactor $\nu$ experiments are needed to shed light on this situation. Nevertheless such experiments can only be conducted at the surface level, where the  backgrounds induced by cosmic rays and 
by the reactor itself are very high. Thus, experiments should provide as much tools as possible to counteract these high backgrounds.

In this context, the SoLid collaboration is operating a 1.6 tons detector at the SCK\raisebox{-0.9ex}{\scalebox{2.8}{$\cdot$}}CEN BR2 research reactor in Belgium. The main goals of SoLid are to confirm or reject the sterile $\nu$ hypothesis as the origin of the RAA, and provide a new measurement of the
$\overline{\nu}_e$'s energy spectrum produced by the $^{235}$U fuel. The imprint induced by a sterile $\nu$  is a distortion in the $\overline{\nu}_e$'s energy spectrum as function of both the $\overline{\nu}_e$ energy and the distance traveled by the $\overline{\nu}_e$'s from the source to the detection point. $\overline{\nu}_e$'s are detected via the inverse beta decay (IBD) process: $\overline{\nu}_e + p \rightarrow e^{+} +n$, where the $\overline{\nu}_e$ energy can be assessed by measuring the $e^{+}$ energy. To look for this signature the SoLid collaboration has developed a novel technology, combining PVT cubes of 5$\times$5$\times$5 cm$^3$ dimensions and $^6$LiF:ZnS(Ag) screens
of $\sim$250 $\mu$m thickness \cite{Abreu:2018pxg}. PVT is used as target for the IBD reaction and at the same time as calorimeter for the $e^{+}$ interaction. On the other hand, $^6$LiF:ZnS(Ag) screens are used for neutron capture and detection. To detect these antineutrino interactions, signals are readout by a network of wavelength shifting fibers and SiPMs.
The SoLid detector consists of 5 modules of 10 planes each, and each plane is made of an array of 16$\times$16 SoLid cells. The fine granularity (12800 cells) provides powerful tools
to distinguish signal from background, but presents a challenge in ensuring homogeneous detector response and calibrating the energy scale and neutron detection efficiency in each cell individually. The physics goals of SoLid requires knowing the energy scale and the neutron detection efficiency in each cell with uncertainties smaller than 2\% and 4\% respectively.

\section{Calibration}
\begin{figure}[h]
\centering
\includegraphics[width=0.35\textwidth,trim={0cm 0.2cm 0cm 0.8cm},clip]{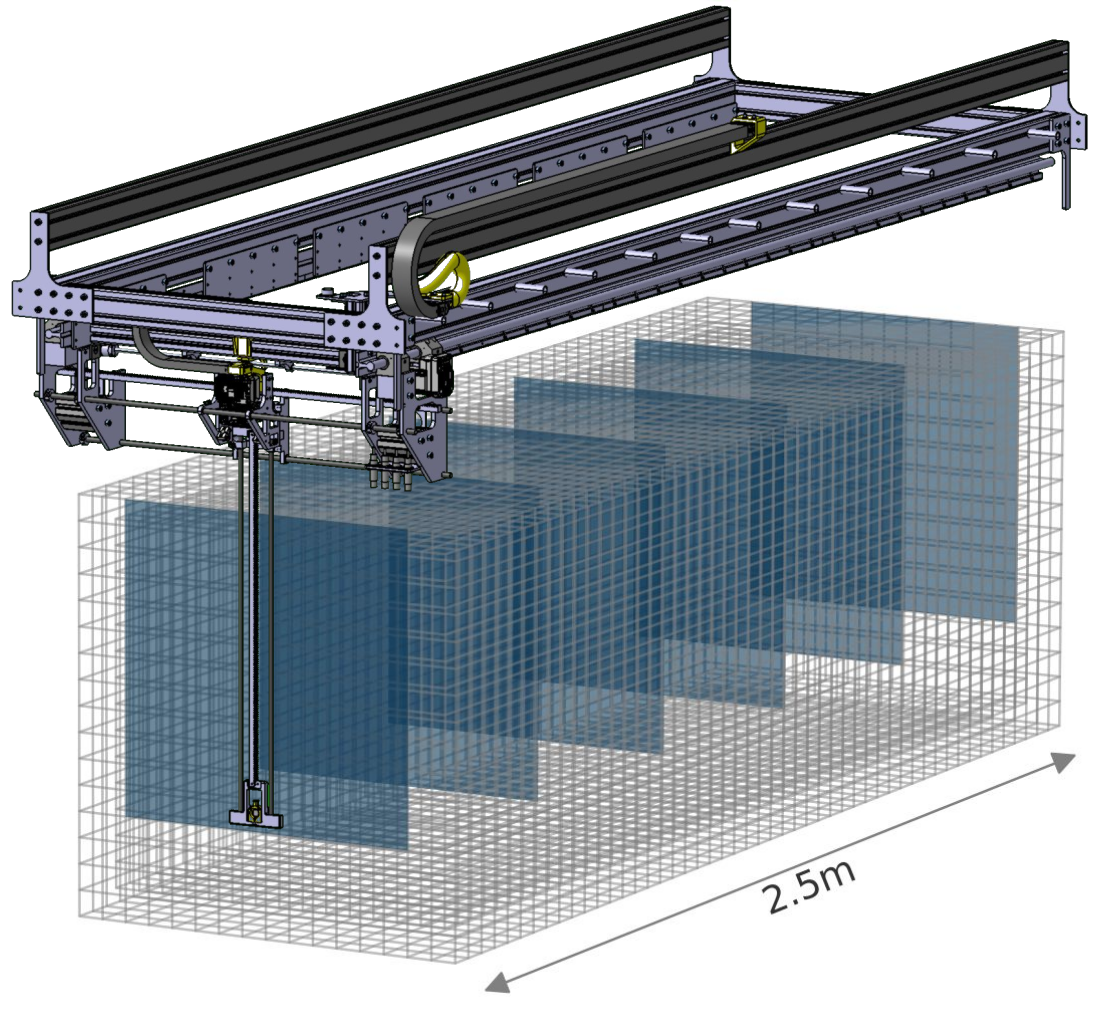}
\caption{CROSS system.}
\label{CROSS}
\end{figure}

To fulfill the SoLid calibration requirements, an automated system called CROSS has been developed. The CROSS system allows to move modules using an actuator system, then radioactive sources are inserted in gaps using a calibration arm as shown in figure \ref{CROSS}.
In this way a maximum distance source-to-cube of less than 35 cm is achieved, which allows for calibrating the full detector. Since the commissioning of the full detector in May of 2018, several calibration campaigns has been successfully completed using different gamma and neutron sources as presented in table \ref{sources}. The gamma and neutron calibration scans of the full detector last one and three days respectively. Periodic calibrations of the full detector are realized every reactor OFF cycle (every two months).

 \begin{center}
  \begin{table}
  \centering
   \begin{tabular}{ c | c | c |c | c | c}
     \hline
     Source & AmBe & $^{252}$Cf & $^{22}$Na & $^{207}$Bi & $^{137}$Cs \\ \hline
     Type & n + $\gamma$ & n & $\gamma$ & $\gamma$ & $\gamma$ \\ \hline
     Energy [MeV] & <4.2> (n) + 4.4 ($\gamma$ ) & <2.1> & 1.27 + 0.511 & 0.57 +1.06 & 0.667\\ 
     \hline
     \end{tabular}
     \caption{Radioactive neutron and gamma sources used for the calibration of the SoLid experiment.}
     \label{sources}
  \end{table}
     \end{center}

\subsection{Energy scale calibration}
Measuring the $\overline{\nu}_e$ energy requires to know precisely the energy scale in each cell individually. Light signals are readout in ADC counts, which are converted to pixel avalanches (PA) taking into account the individual gain of each channel. The PA-to-MeV conversion factor is estimated using gamma sources; a 37 kBq $^{22}$Na source has been selected as the standard source for the calibration of the full detector.   $^{22}$Na provides 1.27 MeV gammas together with 0.511 MeV annihilation gammas. Given the granularity of the detector a full energy peak can not be obtained. Hence, the Compton edge profile of the 1.27 MeV gamma is used for the calibration of the energy scale employing two independent methods \cite{Abreu:2018ekw}. 
\begin{itemize}
\item \textbf{Kolmogorov test:} It consists in smearing a true Geant4 simulated energy spectrum in each PVT cube with a set of energy resolutions and values of light yield. Then a Kolmogorov test between calibration data and each simulated spectrum is realized. The energy scale is taken from the point where the best agreement is found. 
\item \textbf{Analytical fit:}  Use a model based on the Klein-Nishina formula for the cross section of the Compton scattering process. Then it computes the convolution of this cross section and a Gaussian with an energy dependent resolution. The energy scale is estimated comparing the measured Compton edge value (PA) to the predicted one (MeV).
\end{itemize}
Besides the periodic use of the $^{22}$Na source, and for linearity study purposes, dedicated calibration campaigns with other gamma sources have been performed. Results from AmBe and $^{207}$Bi sources show excellent agreement with $^{22}$Na, demonstrating energy linearity in the range of interest for reactor $\overline{\nu}_e$'s.
On top of the dedicated source calibration campaigns, muon tracks are used for a daily monitoring of the energy scale, which shows a very stable detector response.

\subsection{Neutron reconstruction efficiency}
The IBD efficiency is dominated by the neutron detection efficiency. This efficiency is estimated combining the neutron capture efficiency and the reconstruction efficiency: $\epsilon_{det}=\epsilon_{capt}\times\epsilon_{reco}$. The  $\epsilon_{capt}$ is the probability for a neutron to be captured on the $^{6}$Li nuclei, and can be estimated using Geant4 simulations. The 
$\epsilon_{reco}$ is the probability for a neutron capture on the $^{6}$Li to be detected and is the result of the trigger and offline particle identification. In SoLid a dedicated neutron trigger algorithm based on counting the number of Peaks over Threshold (PoT) has been developed \cite{Abreu:2018ekw}.
To assess the neutron detection efficiency, AmBe and $^{252}$Cf neutron sources are used, which allows to decrease systematic uncertainties. 

\begin{figure}[h]
\centering
\includegraphics[width=0.7\textwidth,trim={0cm 0.2cm 0cm 0.8cm},clip]{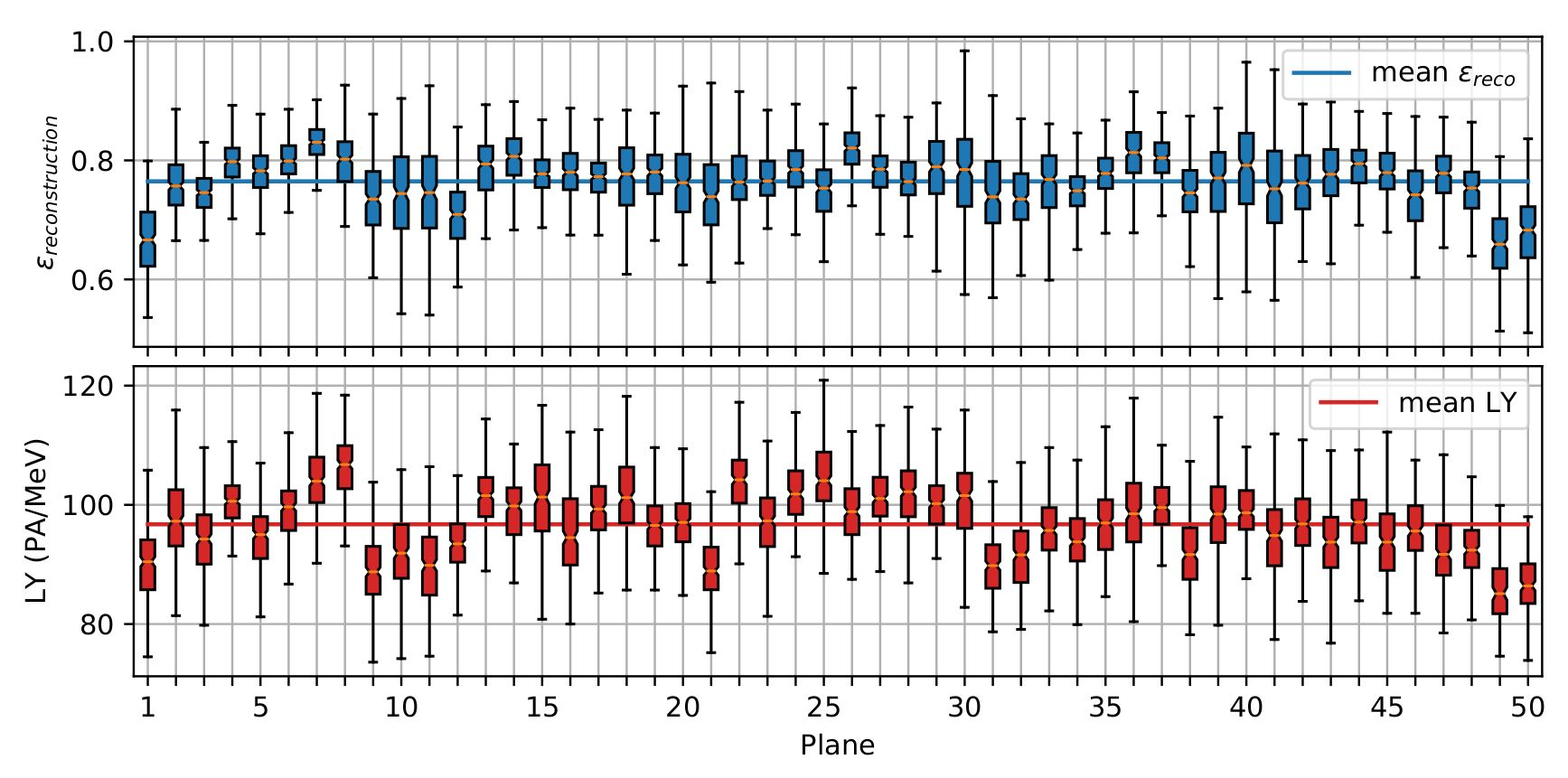}
\caption{Light yield and neutron reconstruction efficiency of the 50 planes of the SoLid detector.}
\label{CalibResults}
\end{figure}

\section{Conclusions}
The SoLid collaboration has developed a new detector concept for reactor $\overline{\nu}_e$ detection using a highly segmented detector.
We have developed methods in order to calibrate each cell of such a segmented detector. Thus the light yield and the neutron reconstruction efficiency have been measured in all the 12800 detector cells. 
On average a light yield of 77 PA/MeV with a stochastic resolution of 12\% at 1 MeV has been measured. In addition, excellent agreement between different radioactive sources at different energies proves energy linearity of PVT, which is also confirmed with muon calibration at high energy. On the other hand, a
neutron reconstruction efficiency of about 75 \% has been measured, which will result in a good IBD detection efficiency.
The calibration results show a good detector homogeneity, with a dispersion smaller than 10\% for both light yield and neutron detection efficiency

\end{document}